\documentclass[aip,jap,twocolumn,reprint]{revtex4-1}
\usepackage{amsmath}
\usepackage{amssymb}
\usepackage{graphicx}
\usepackage{hyperref}
\begin{document}
\title{Magnetic orders of LaTiO$_3$ under epitaxial strain: a first-principles study}
\author{Yakui Weng}
\author{Xin Huang}
\author{Yankun Tang}
\author{Shuai Dong}
\email{sdong@seu.edu.cn}
\affiliation{Department of Physics, Southeast University, Nanjing 211189, China}
\date{\today}

\begin{abstract}
Perovskite LaTiO$_3$ bulk is a typical Mott-insulator with G-type
antiferromagnetic order. In this work, the biaxial strain effects on the ground
magnetic order of LaTiO$_3$ films grown on various substrates have been studied.
For the compressive strain, LaTiO$_3$ films grown on LaAlO$_3$, LaGaO$_3$, and
SrTiO$_3$ substrates undergo a phase transition from the original G-type
antiferromagnet to A-type antiferromagnet. The underlying physical mechanisms
are the lattice distortions tunned by strain. While for the tensile strain, the
BaTiO$_3$ and LaScO$_3$ substrates have been tested, which show a tendency to
transit the LaTiO$_3$ to the C-type antiferromagnet. Furthermore, our
calculations find that the magnetic transitions under epitaxial strain do not
change the insulating fact of LaTiO$_3$.
\end{abstract}
\maketitle

\begin{figure}
\centering
\includegraphics[width=0.5\textwidth]{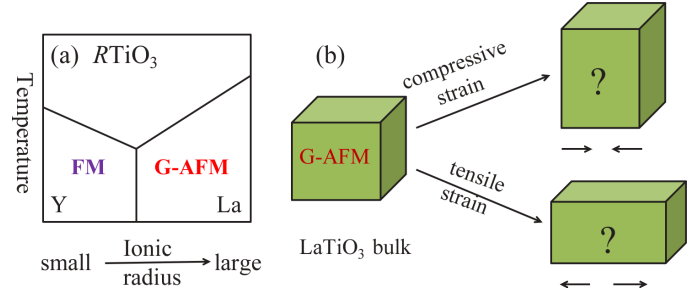}
\caption{(a) Sketch of experimental magnetic phase diagram of $R$TiO$_3$. (b)
Sketch of our motivation: LaTiO$_3$ films grown on various substrates under
compressive or tensile strain.}
\label{diagram}
\end{figure}

Perovskite oxides $AB$O$_3$ have attracted continuing attention and been
intensively investigated due to their novel physical properties and a broad
range of technical applications.\cite{Dagotto:Sci} Among abundant
perovskite compounds, the canonical Mott insulator $R$TiO$_3$ ($R^{3+}$ denotes
a rare-earth ions) are physically interesting due to the complex couplings
between the orbital, spin, lattice degrees of freedom of Ti's $3d$ electron
which is localized by the strong Coulombic interaction. In $R$TiO$_3$
perovskites, the
ligand crystal field from the oxygen octahedron splits the $5$-fold $3d$ levels
into two groups: the $3$-fold $t_{\rm 2g}$ orbitals and the $2$-fold $e_{\rm g}$
orbitals. The Fermi level is located in the $t_{\rm 2g}$ levels, and the $t_{\rm
2g}$ orbitals are highly localized due to the $p$-$d$ hybridization. Moreover,
the
GdFeO$_3$-type structure distortions, which combine the tilts and rotations of
the oxygen octahedrons, are prominent in the orthorhombic $R$TiO$_3$. According
to previous studies,\cite{Imada:Rmp,Solovyev:Prb,Okatov:Epl} the
ground magnetic phase of $R$TiO$_3$ transits from the ferromagnetism to G-type
antiferromagnetism, with increasing size of $R$ or in other words with weaking
GdFeO$_3$-type distortions,\cite{Mochizuki:Njp} as shown in
Fig.~\ref{diagram}(a).

In the $R$TiO$_3$ bulks, compounds with small GdFeO$_3$-type distortions exhibit
the G-type antiferromagnetic (AFM) ordering, e.g. LaTiO$_3$, and the large
ones tend to lead the FM
ordering, e.g. YTiO$_3$. In addition, it is well known that perovskite oxides
may be sensitive to external factors.\cite{Dong:Prb,Dong:Prb12,Bousquet:Prb} For
example, recently the use of epitaxial strain has attracted great attentions due
to many unexpected effects on thin
films,\cite{Rondinelli:Am,Bousquet:Prb,Schlom:Armr,Zubko:Arcmp,Dong:Prb08.3}
which has been proved
to be a useful route to design potential devices.

In this work, the effects of epitaxial strain on the ground magnetic order  of
LaTiO$_3$ films will be studied using the first-principles calculations, as
illustrated in Fig.~\ref{diagram}(b). Our
calculations predict that a robust A-type AFM phase can be stabilized by the
compressive strain. In contrast, LaTiO$_3$ films remain G-type AFM
under moderate tensile strain, but have a tendency to become the C-type AFM with
further increasing the tensile strain.

LaTiO$_3$ has the orthorhombic structure (space group
\textit{Pbnm})\cite{Rondinelli:Am,Cwik:Prb} with
the experimental lattice constants of $a$=$5.636$ \AA{}, $b$=$5.618$ \AA{}, and
$c$=$7.916$ \AA{}, containing 4 formula units.\cite{Komarek:Prb} In the
following, two different strain have been considered: in-plane compressive vs
tensile. Five widely used substrates have been tested, including LaAlO$_3$
($\sqrt{2}a$=$\sqrt{2}b$=$5.366$ \AA{}), LaGaO$_3$ ($a=5.49$ \AA{}, $b$=$5.53$
\AA{}), SrTiO$_3$ ($\sqrt{2}a$=$\sqrt{2}b$=$5.523$ \AA{}) for the compressive
case and BaTiO$_3$ ($\sqrt{2}a$=$\sqrt{2}b$=$5.65$ \AA{}), LaScO$_3$
($a$=$5.678$ \AA{}, $b$=$5.787$ \AA{}) for the tensile case. Here LaTiO$_3$ is
assumed to be grown along the
most studied ($001$) direction. Our first-principles density-functional theory (DFT) calculations are performed using the local density approximation (LDA)
method with the Hubbard $U$ and the projector-augmented wave (PAW) potentials, as
implemented in the Vienna \textit{ab} initio Simulation Package
(VASP).\cite{Kresse:Prb,Kresse:Prb96} The on-site Hubbard interaction is set as
$U-J=2.3$ eV using the Dudarev implementation\cite{Dudarev:Prb} for the
localized $3d$ electrons of Ti. The lattice constants are fixed to match the
particular substrate. Then the lattice constant along the ($001$) direction and
inner atomic positions are fully optimized as the Hellman-Feynman forces are
converged to less than $1.0$ meV/\AA{}. The cutoff energy of plane-wave is
$500$ eV and the Brillouin-zone integrations are performed with the tetrahedron
method\cite{Blochl:Prb} over a $7\times7\times5$ Monkhorst-Pack
$k$-point\cite{Monkhorst:Prb} mesh centered at $\Gamma$.

\begin{figure}
\centering
\includegraphics[width=0.5\textwidth]{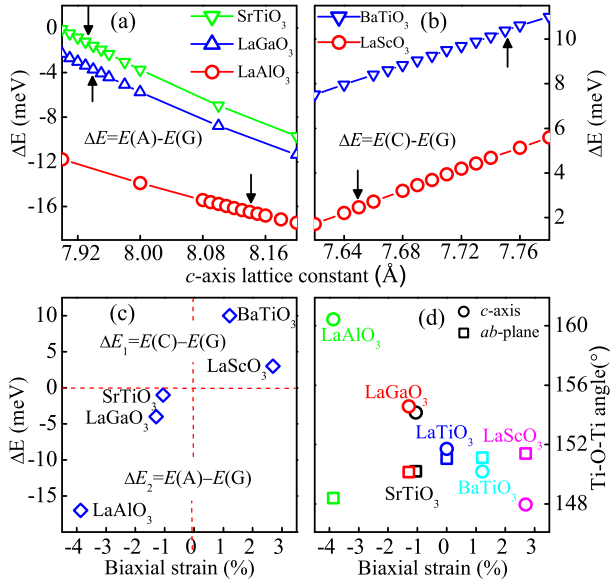}
\caption{(Color online) (a-b) The energy difference per Ti as a function of $c$
lattice constant. (a) Between the A-AFM and G-AFM for the compressive cases.
(b) Between the C-AFM and G-AFM for the tensile cases.
(c) The energy difference (per Ti) on different substrates as a function of the biaxial strain. For the tensile strain, $\Delta
E_1$=$E$(C-AFM)-$E$(G-AFM) and for the tensile
strain, $\Delta E_2$=$E$(A-AFM)-$E$(G-AFM).
(d) The Ti-O-Ti bond angles (both in-plane and out-of-plane) as a function of
strain.}
\label{energy}
\end{figure}

\begin{figure}
\centering
\includegraphics[width=0.5\textwidth]{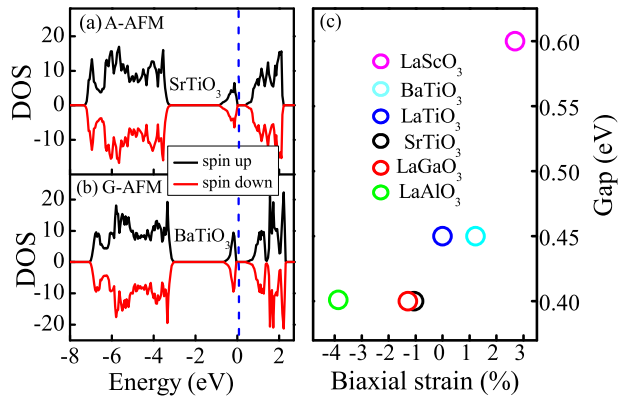}
\caption{(Color online) Total DOSs of LaTiO$_3$ films under strain. (a) on
SrTiO$_3$. (b) on BaTiO$_3$. The Fermi energy is located in zero. (c) The
energy gap as a function of strain.}
\label{DOS}
\end{figure}

First, the ground state of bulk LaTiO$_3$ has been checked. The lattice is
fully optimized, giving $a$=$5.615$ \AA{}, $b$=$5.549$ \AA{}, and $c$=$7.828$ \AA{}
which are close to the experimental data. The non-magnetic (NM) state and four
magnetic orders: ferromagnetic (FM), A-type AFM, C-type AFM, and G-type AFM,
are calculated and compared in energy. As shown in Table I, the G-type AFM is the most stable
state, as found in experiments. The calculated magnetic moment is
$0.75$ $\mu_B$/per Ti, slightly larger than the experimental result
$0.57$ $\mu_B$.\cite{Cwik:Prb} Our DFT calculations (Fig.~\ref{DOS}(c)) find
the insulating behavior with an energy gap of $0.45$ eV, in agreement with previous
DFT result\cite{Okatov:Epl} and a little overestimated compared with experimental value $0.2$
eV,\cite{Okimoto:Prb} implying a Mott-insulator.

\begin{table}
\caption{The energy difference and corresponding magnetic moment per Ti in unit
of $\mu_B$ of unstrained bulk LaTiO$_3$: $\Delta E$ (per
Ti)=$E$(magnetic)-$E$(FM).}

\begin{tabular*}{0.48\textwidth}{@{\extracolsep{\fill}}llllllr}
\hline \hline
Magnetic order & NM & FM & A-AFM & C-AFM & G-AFM \\
\hline
$\Delta E$ & $124$  & $0$  & $-13$  & $17$  & $-18$\\
Magnetic moment &  $0$  & $0.86$  & $0.80$  & $0.77$  & $0.75$\\
\hline \hline
\end{tabular*}
\label{table}
\end{table}

Subsequently, the effects of strain will be studied. For the compressive
strain, the small lattice SrTiO$_3$, LaGaO$_3$, and LaAlO$_3$ substrates are adopted as
the weak, middle, and strong cases. The internal atomic positions are relaxed
with various magnetic orders within a wide range from $7.6$ \AA{} to $8.6$
\AA{} for lattice constant along the $c$-axis to search the optimized structure and
ground state. The obtained equilibrium values for the $c$-axis are around
$7.93$\AA{}, $7.94$ \AA{}, and $8.14$ \AA{} for SrTiO$_3$, LaGaO$_3$, and LaAlO$_3$,
respectively. In all these cases, the total energies show that A-type AFM is
the most stable state with the relaxed structure, instead of the G-type AFM in
bulk. Moreover, the FM and C-type AFM are much higher in energy than the A- and
G-type AFMs. Therefore, in the following, only the results of A- and G-type AFMs will
be presented for the compressive substrates.

The energy differences between these two orders are showed in Fig.~\ref{energy}(a). The A-type AFM is
most robust ($17$ meV/Ti lower in energy) when grown on the LaAlO$_3$ substrate
with the smallest in-plane lattice, while it is very fragile (only $1$ meV/Ti
lower in energy) on SrTiO$_3$. As shown in Fig.~\ref{energy}(c), epitaxial LaTiO$_3$ films
on these three substrates would have a biaxial compression of about $\sim3.8\%$
for LaAlO$_3$, $\sim1.3\%$ for LaGaO$_3$, and $\sim1.0\%$ for SrTiO$_3$, suggesting
the direct relation between the magnetism and strain. In fact, our previous
calculation also predicted the A-type AFM state appeared in the YTiO$_3$ film
on the ($001$) LaAlO$_3$ substrate which is FM in bulk.\cite{Huang:Jap} The
A-type AFM state does not exist in any $R$TiO$_3$ bulk so far, but may be obtained in compressive
films despite the original states (FM or G-type AFM).

Next, the tensile strain effects will be studied in the same way, using
BaTiO$_3$ and LaScO$_3$ as the substrates. The relaxed lattice constant along
the $c$-axis is about $7.65$ \AA{} for LaScO$_3$ substrate and $7.75$ \AA{} for
BaTiO$_3$ substrate. Different from the strain-driven phase transition in
compressive cases, LaTiO$_3$ films remain in the G-type AFM order as in the
bulk. In the tensile case, the FM and A-type AFM states have relatively higher
energies than the G- and C-type AFM ones which are very proximate in energy. As
shown in Fig.~\ref{energy}(b), with decreasing length of $c$-axis, the energy differences
between the G- and C-type AFMs decease, e.g. $10$ meV/Ti for equilibrium length
on the BaTiO$_3$ substrate, and $3$ meV/Ti for the LaScO$_3$ case. These
results show that the tensile LaTiO$_3$ films have an obvious tendency to be C-type AFM
if further large lattice substrates are used. These new phases (the A-type and
possible C-type AFMs) are physical interesting, which enrich the magnetic phase
diagram of the $R$TiO$_3$ family.

As stated before, the Ti-O-Ti bond angles can be used as a parameter to
characterize the lattice distortions in $R$TiO$_3$. As shown in Fig.~\ref{energy}(d),
with increasing biaxial compression, the bond angles decrease in the $ab$-plane but
increase along the $c$-axis, while the tensile strain gives the opposite trend.
According to $R$TiO$_3$  bulk's phase diagram, the relation between lattice
distortions and magnetic orders are well-established: FM order for strong
distortions (small Ti-O-Ti bond angles), AFM order for weak distortions (large
Ti-O-Ti bond angles). Thus, the compressive strain, which decreases the
in-plane bond angles but increase the out-of-plane one, tends to make spins
arrange parallel in-plane and anti-parallel along the $c$-axis, namely the
A-type AFM order. In contrast, for the tensile cases, the opposite changes of
bond angles favor the AFM coupling in-plane but FM coupling along the $c$-axis,
namely the C-type AFM tendency although it has not be achieved on BaTiO$_3$
and LaScO$_3$
substrates.

Moreover, for all strained LaTiO$_3$, the insulating behavior has been
preserved despite the magnetic phase transitions. For example, the DOSs of LaTiO$_3$
films grown on SrTiO$_3$ and BaTiO$_3$ substrates are shown in Fig.~\ref{DOS}(a) and ~\ref{DOS}(b),
respectively. In both cases, a gap exists at the Fermi level. The states near
the Fermi level is dominated by Ti $t_{\rm 2g}$ levels. The tiny difference of
DOSs between the SrTiO$_3$ and BaTiO$_3$, can also reflect the strain effect to
the
electronic structure. As summarized in Fig.~\ref{DOS}(c), the band gap increases
slightly from the compressive strain to tensile strain.

In conclusion, the magnetic orders of LaTiO$_3$ films with the biaxial
compressive and tensile strain have been studied using LDA+$U$ method. For the
compressive strain, a phase transition from G-type AFM to A-type AFM has been
found and this transition is much more robust when the strain increases.
However, the G-type AFM still be the ground state for the tensile strain and the C-type
AFM maybe appear if the strain is further increased. Furthermore,
the LaTiO$_3$ films preserve the insulating behavior on all substrates studied
here.

Work was supported by the 973 Projects of China (Grant No. 2011CB922101), NSFC
(Grant Nos. 11274060, 51322206).

\end{document}